\documentclass[letter,12pt]{article}
\usepackage{graphicx}

\newcommand{\form}{\mbox{C$_{12}$EO$_{6} \,$}}
\newcommand{\formeau}{\mbox{C$_{12}$EO$_{6}$/H$_{2}$O$\,$}}
\newcommand{\dgr}{ {\,}^{\circ} \mbox{C}}
\newcommand{\un}[1]{\ensuremath{\unskip\,\mathrm{#1}}}

\begin{document}

\date{}
\title{Structural Transition in the Isotropic Phase of the {\formeau} Lyotropic Mixture~: A Rheological Investigation}

\author{D. Constantin\protect\footnote{Author for correspondence. Present address~: Institut f\"{u}r R\"{o}ntgenphysik,
Geiststra{\ss}e 11, 37073 G\"{o}ttingen. E-mail~:
dcconsta@ens-lyon.fr; Tel : +49-551 39 50 66; Fax : +49-551 39 94
30.} {}, \'{E}. Freyssingeas, J.-F. Palierne  and P. Oswald
\\ {\em Laboratoire de Physique de l'ENS de Lyon,} \\ {\em 46
All\'ee d'Italie, 69364 Lyon Cedex 07, France}}

\maketitle

\begin{abstract}
\noindent We study the structural changes occurring in the
isotropic phase of the {\formeau} lyotropic mixture (up to $35 \%$
surfactant weight concentration) upon increasing the concentration
and temperature, from small individual micelles to an entangled
network which subsequently becomes connected. High-frequency (up
to $\omega = 6 \,10^4 \un{rad/s}$) rheological measurements give
us access to the viscoelastic relaxation spectrum, which can be
well described by the sum of two Maxwell models with very
different temperature behaviors~: the slower one ($\tau _1 \sim
10^{-4} \un{s}$) is probably due to reptation and its associated
viscosity first increases with temperature (micellar growth) and
then decreases after reaching a maximum (appearance of
connections). The fast mechanism ($\tau _2 \sim 10^{-6} \un{s}$)
remains practically unchanged in temperature and can be related to
the relaxation of local micellar order, as observed at higher
concentration in a previous investigation. This interpretation is
confirmed by additional measurements in aqueous mixtures of the
related surfactant \mbox{C$_{12}$EO$_{8} \,$} (which forms smaller
micelles), where only the fast mechanism --related to local
order-- is detected.

\end{abstract}

PACS~: 61.30.St, 82.70.Uv, 83.80.Qr


\section{Introduction}

The isotropic phase of the binary system {\formeau}, as well as
those formed by similar non-ionic surfactant molecules, have been
studied for more than twenty years; in the beginning, research
focused on the structure at low surfactant concentration,
especially on the temperature evolution of micellar size and
shape. It is now well established that the micelles have a general
tendency of increasing in size and becoming anisotropic
(cylindrical) with increasing temperature and concentration (see
\cite{lindman91} and references therein). It has also been shown
that micellar growth strongly depends on the molecular details~:
it is very important for  \mbox{C$_{12}$EO$_{5} \,$} and {\form}
but much more modest for \mbox{C$_{12}$EO$_{8} \,$}, which forms
short micelles.

When they are not too long, these elongated micelles can be seen
as rigid rods, but once they exceed a certain persistence length
$\ell _{\rm p}$, they become flexible (estimations for the
persistence length of {\form} vary from 7 nm \cite{ravey83} to 25
nm \cite{carale92}). If the micelles are much longer than $\ell
_{\rm p}$ (``wormlike micelles''), they assume very complex
configurations resembling polymers in solution, with the essential
difference that the micellar length is not chemically fixed, but
rather fluctuates around an equilibrium length depending on the
thermodynamical parameters, hence the name of ``living polymers''.
Above an overlap concentration $c_{\rm c}$, the micelles begin to
touch and become entangled. A quantitative estimate of the curve
$c_{\rm c}(T)$ for the {\formeau} system was given in reference
\cite{carale92}.

The rheology of entangled polymers is well established
\cite{ferry80,larson99}, and its concepts were recently applied to
wormlike micellar systems \cite{cates90,granek92,granek94}, taking
into account the dynamical nature of the micellar length
distribution (see \cite{cates96} for a review). This model was
successfully used to describe the rheological behavior of micellar
solutions of ionic surfactants \cite{lequeux94}.

Another interesting property of wormlike micelles is their
tendency of interconnecting under certain conditions
(concentration, temperature or -in ionic systems- counterion
concentration). Such behavior, first found in ternary systems
\cite{porte86,ninham87}, was subsequently observed in a very wide
range of binary mixtures of single-tail non-ionic
\cite{alami93,kato94,kato95,kato96,mallamace95,bernheim00} or
double-tail zwitterionic \cite{shchipunov98,ambrosone01}
surfactants, as well as in pseudo-binary systems (ionic surfactant
+ brine)
\cite{monduzzi93,appell92,khatory93b,narayanan97,hassan98,aitali97,aitali99,raghavan02}.

Structurally, connected and entangled systems only differ on a
very small scale, which makes them difficult to tell apart using
static techniques. Furthermore, the microscopic structure of the
connections is far from being elucidated (see \cite{may97} and
references therein). Indirect techniques must then be employed;
for instance, the minimum in surfactant self-diffusion coefficient
as a function of the concentration appearing in a wide variety of
systems was often explained by the appearance of connections
\cite{kato94}. However, it was shown \cite{schmitt94} that this
behavior could also originate in the competition between two
different mechanisms~: diffusion of the micelle itself and
diffusion of the surfactant molecule on the micelle.

On the other hand, connections can have a dramatic effect on the
{\em dynamical} properties of wormlike micellar systems. Contrary
to connected polymer systems, where the reticulation points
(permanent chemical cross-links) slow down the dynamics,
connections between entangled micelles can actually facilitate
relaxation and render the system more fluid. This aspect prompted
a systematic study of the connectivity in micellar solutions~:
several experiments \cite{appell92,khatory93b,rehage88} showed
that, in ionic surfactant systems, the viscosity decreases on
increasing the salt concentration. They were followed by
theoretical works on the conditions of connection formation
\cite{drye92} and their influence on the dynamics of the phase
\cite{lequeux92}. The theory was then employed to qualitatively
characterize the appearance of connections
\cite{narayanan97,hassan98,aitali99}. Briefly, branching points
allow the surfactant to ``flow'' more easily across the micellar
network, thus increasing the curvilinear diffusion constant of a
micelle $D_{\rm{c}}$. As the terminal relaxation time in entangled
systems is the reptation time $\tau_{\rm R} = L^2/D_{\rm{c}}$,
this amounts (from the dynamical point of view) to replacing the
average length of a micelle by the average distance between
branching points along a micelle \cite{cates96}. When the latter
becomes of the order of the entanglement length, the network is
termed ``saturated'' \cite{drye92}; in this case, the concept of
reptation is no longer valid and other relaxation mechanisms, such
as those related to the local order, can become dominant
\cite{constantin02}.

The purpose of this article is to study the structural changes
that occur in the isotropic phase of an aqueous solution of
non-ionic surfactant upon increasing the concentration or the
temperature, from small individual micelles to an entangled
network which then becomes connected. We achieve this by relating
the viscoelastic relaxation mechanisms to the structural features
and by monitoring their evolution with concentration and
temperature.

We employ high-frequency rheology to study the isotropic phase in
the {\formeau} lyotropic mixture, where {\form} is the non-ionic
sur\-fac\-tant hexa-ethylene glycol mono-n-dodecyl-ether, or
\ensuremath{\mathrm{CH_3(CH_2)_{11}(OCH_2CH_2)_6OH}} (for the
phase diagram see \cite{mitchell83}). Its dynamic behavior has
already been investigated by measuring the shear viscosity
\cite{strey96,darrigo98}, sound velocity and ultrasonic absorption
\cite{darrigo98}, as well as NMR relaxation rates
\cite{burnell00}, all pointing to the presence of wormlike
micelles (at least above 10 \% surfactant concentration by weight
\cite{darrigo98}). In previous experiments
\cite{sallen96,constantin01} we have shown that, for 50 \% wt
surfactant concentration, above the hexagonal mesophase, the
isotropic phase has a structure consisting of surfactant cylinders
that locally preserve the hexagonal order over a distance $d$
varying from about $40 \, \un{nm}$ at $40 \dgr$ to $25 \, \un{nm}$
at $60 \dgr$. Local order is equally evidenced at lower
concentration (down to about 20\%) by SANS measurements
\cite{zulauf85} (let us mention that in some systems local order
can appear at very low concentrations, as shown by K\'{e}kicheff
{\em et al.} \cite{kekicheff94}). Between the cylinders there is a
large number of thermally activated connections (with an estimated
density $n \sim 10^{6} \, \un{\mu m^{-3}}$) \cite{constantin01}.
Investigating the system at lower concentration, over a wide
temperature range, allows us to describe its evolution from small
individual micelles to a completely connected network.

For comparison, we also study aqueous solutions of the related
surfactant \mbox{C$_{12}$EO$_{8} \,$}, which is known to form
smaller micelles \cite{lindman91}. The difference in rheological
behavior between the two systems is discussed in terms of the
structure.

\section{Materials and Methods}

The surfactants were purchased from Nikko Chemicals Ltd. and used
without further purification. We used ultrapure water ($\rho = 18
\un{M \Omega \, cm}$) from the in-house ELGA system for the
{\formeau} system. The \mbox{C$_{12}$EO$_{8} \,$}, on the other
hand, was mixed with $\un{D_{2}O}$, as the samples were also used
for neutron scattering experiments. The samples were prepared by
weighing the components directly into the vials. The mixtures were
carefully homogenized by repeatedly heating, stirring and
centrifuging and then allowed to equilibrate at room temperature
over a few days.

Rheology measurements were performed in a piezorheometer, the
principle of which has been described in reference
\cite{cagnon80}~: the liquid sample of thickness $40$ or $60
\un{\mu m}$ is contained between two glass plates mounted on
piezoelectric ceramics. One of the plates is made to oscillate
vertically with an amplitude of about 1 nm by applying a sine wave
to the ceramic. This movement induces a squeezing flow in the
sample and the stress transmitted to the second plate  is measured
by the other piezoelectric element. The shear is extremely small~:
$\gamma \leq 2 \, 10^{-3}$, so the sample structure is not altered by
the flow. The setup allows us to measure the storage ($G'$) and
loss ($G''$) shear moduli for frequencies ranging from $1$ to $6
\, 10^{4} \un{rad/s}$ with five points per frequency decade. The
maximum shear rate is thus $\dot{\gamma} \sim 120 \un{rad/s}$, much
lower than the characteristic relaxation times (see section 4). The
entire setup is temperature regulated within $0.05 \dgr$ and
hermetically sealed to avoid evaporation.

\section{Theoretical Considerations}

The main relaxation process in entangled polymer solutions is
reptation, by which the chain gradually disengages from its
initial deformed environment (``tube'') and adopts a stress-free
configuration. The typical reptation time is given by~: $\tau
_{\rm{rep}} \simeq L ^2 / D_{\rm{c}}$, with $L$ the chain length
and $D_{\rm{c}}$ the curvilinear diffusion constant of the micelle
in its tube \cite{larson99}, and the shear modulus exhibits
exponential decay~: $G(t) \sim \exp (-t/\tau _{\rm{rep}})$. In the
case of wormlike micellar systems, the length distribution $c(L)$
must be taken into account and the resulting relaxation is highly
non-exponential \cite{cates90}~:
\begin{equation}
G(t) \sim \exp \left [ - \left ( \frac{t}{\tau _{\rm{rep}}} \right )^{1/4} \right ] \, ,
\label{eq:grep}
\end{equation}
where $\tau _{\rm{rep}}$ is now given by~:
\begin{equation}
\tau _{\rm{rep}}=\frac{L_{\rm{m}} ^2}{D_{\rm{c}}}\, ,
\label{eq:taurep}
\end{equation}
with $L_{\rm{m}}$ the micellar length averaged over $c(L)$.

Another distinctive feature of wormlike micelles is that they can
break up and reform, with a typical time $\tau _{\rm{br}}$. For
$\tau _{\rm{br}} \gg \tau _{\rm{rep}}$, the dynamical response
(\ref{eq:grep}) is not affected; in the opposite limit, $\tau
_{\rm{br}} \ll \tau _{\rm{rep}}$, $G(t)$ approaches a single
exponential \cite{cates90}, with relaxation time $\displaystyle
\tau \simeq \sqrt{\tau _{\rm{br}} \tau _{\rm{rep}}}$. Besides
reptation, which is a process involving the entire chain,
additional (local) relaxation modes are present at higher
frequency, given by ``breathing'' (tube length fluctuations) and
Rouse dynamics \cite{granek92,granek94}.

\begin{figure}[htbp]
\centerline{\includegraphics[width=12cm]{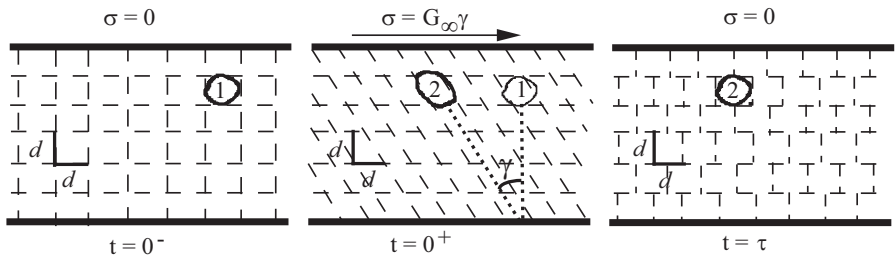}}
\caption{\protect\small Schematic representation of shear in a
material consisting of units of typical size $d$. One such unit
(thick line contour) has been displaced between points 1 and 2.
The instantaneous elastic stress is $\sigma = G_{\infty} \gamma$;
it relaxes over a typical time $\tau$ given by equation
(\ref{tau}).} \label{shear}
\end{figure}

In a previous paper \cite{constantin02} we showed that in the same
mixture at higher concentration, above the hexagonal phase
($c=50\%$, $T>40 \dgr$), the isotropic phase is highly connected
and the viscoelastic response is given by the relaxation of the
local micellar order evidenced by X-ray scattering
\cite{constantin01}. As discussed in reference
\cite{constantin02}, when the system exhibits local order (induced
by the micelle-micelle interaction) with a range $d$, one can only
observe elastic behavior by probing the system on scales smaller
than this correlation distance. The time $\tau$ needed to relax
the stress can then be estimated as~:
\begin{equation}
\label{tau}
 \tau  \sim \frac{d^2}{6 D} \, ,
\end{equation}
where $D$ is the collective diffusion constant. A pictorial representation is given in figure \ref{shear}~: consider a
material with short-range order confined between two plates. The system can be seen as consisting of
elasticity-endowed units of typical size $d$, the correlation distance. After applying an instantaneous shear $\gamma$
by moving the upper plate to the left, one such unit (represented in thick line) has been advected from point 1 to
point 2. At time $t=0^+$ after the deformation, the stress on the upper plate is $\sigma = G_{\infty} \gamma$, with
$G_{\infty}$ the instantaneous (``infinite'' frequency) shear modulus of the elastic material. Since there is no
long-range restoring force, once the particles equilibrate their internal configuration (over a distance $d$), the
elastic stress is completely relaxed; thus, after a time $\tau$ given by eq. (\ref{tau}), $\sigma = 0$. At $c=50\%$, $d
\sim 25-40 \un{nm}$ and $D \sim 1.5 \, 10^{-10} \un{m^2/s}$, leading to a relaxation time $\tau \sim 1 \un{\mu s}$.
$D$ decreases at lower concentration \cite{brown87}, as well as the correlation range \cite{zulauf85}, but we have no
quantitative estimate for the latter. For the plateau modulus $G_{\infty}$ one can take the shear modulus of the
hexagonal phase, because at high frequency the structure is probed on scales smaller than the correlation length,
where it is locally organized. Its value, $G_{\infty} \sim 5 \, 10^4 \un{Pa}$, is in good agreement with the
experimental results. The modulus varies rapidly with the distance $a$ between micelles ($G_{\infty} \sim k_B T / a^3
$), so it should decrease at lower concentration.

When the system is not completely connected, both previously described processes are relevant so, in the simplest
approach, we can assume that the shear modulus is a sum of two mechanisms, one related to polymer-like dynamics and the
other given by order relaxation, each one with a characteristic time scale. Thus, in the first approximation we expect
a bimodal relaxation of the form $G(t) = G_{\infty 1} \exp (-t/\tau _{1}) + G_{\infty 2} \exp (-t/\tau _{2})$ or, in
the frequency domain~:
\begin{equation}
G^*(\omega) = G' + i G'' = \frac{i \omega \eta _1}{1+i \omega
\tau_1}+\frac{i \omega \eta _2}{1+i \omega \tau_2} \, ,
\label{eq:2maxwell}
\end{equation}
where $\tau_1$ and $\tau_2$ are the respective relaxation times,
while $G'$ and $G''$ are the storage and loss moduli, describing
elasticity and dissipation, respectively. We express the complex
shear modulus $G^*(\omega)$ as a function of the terminal
viscosities $\eta _1$ and $\eta _2$ which can be more reliably
determined from the low-frequency data than the plateau moduli
$G_{\infty i} = \eta_{i} / \tau_{i} \, , \, i=1,2$. For
definiteness, subscript `1' will denote the slower relaxation
mechanism ({\it i. e.} $\tau_{1} >\tau_{2}$).

\section{Results and Discussion}

\begin{figure}[htbp]
\centerline{\includegraphics[width=17.5cm]{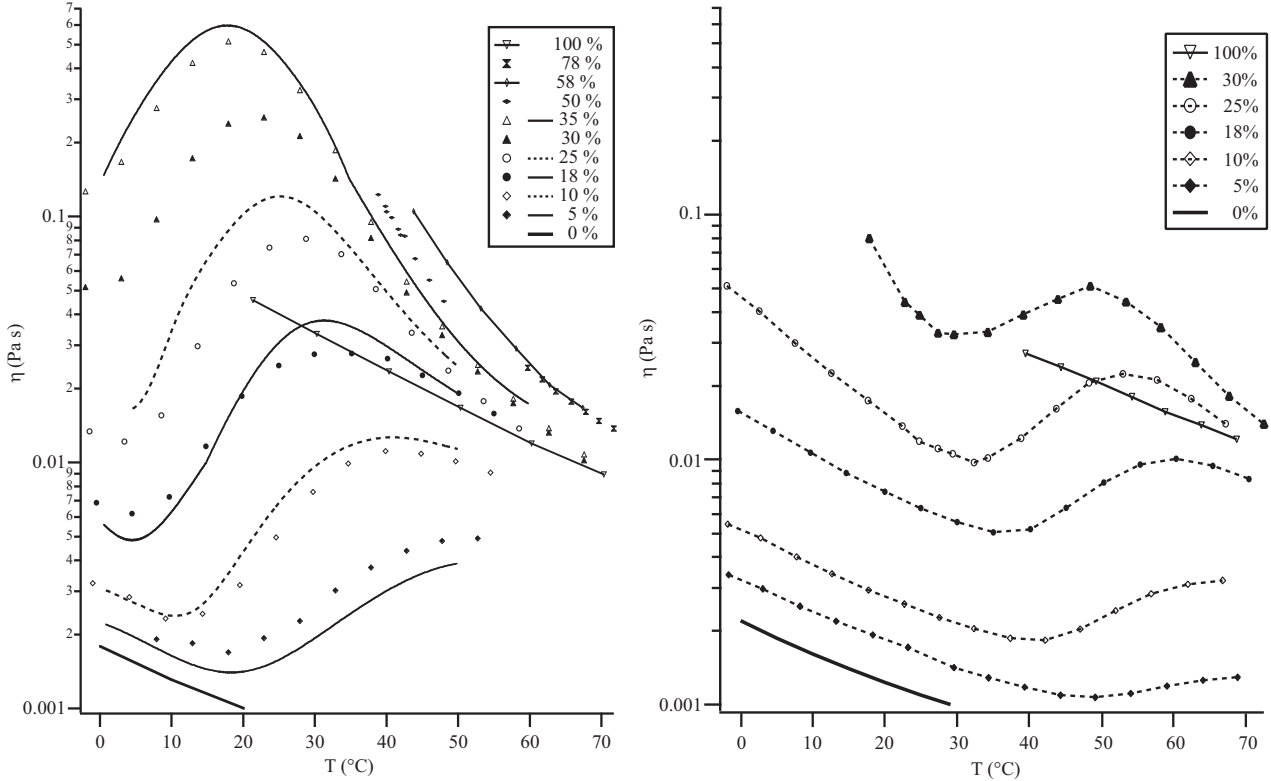}}
\caption{\protect\small Left~: Low frequency viscosity at
different concentrations and temperatures in the isotropic phase
of the {\formeau} mixture (various symbols) compared to the
results of Strey \cite{strey96} (full and dotted lines). Right~:
Low frequency viscosity in the isotropic phase of the
\mbox{C$_{12}$EO$_{8}$/D$_{2}$O$\,$} mixture.}
\label{fig:eta_total}
\end{figure}

\begin{figure}[htbp]
\centerline{\includegraphics[width=17.5cm]{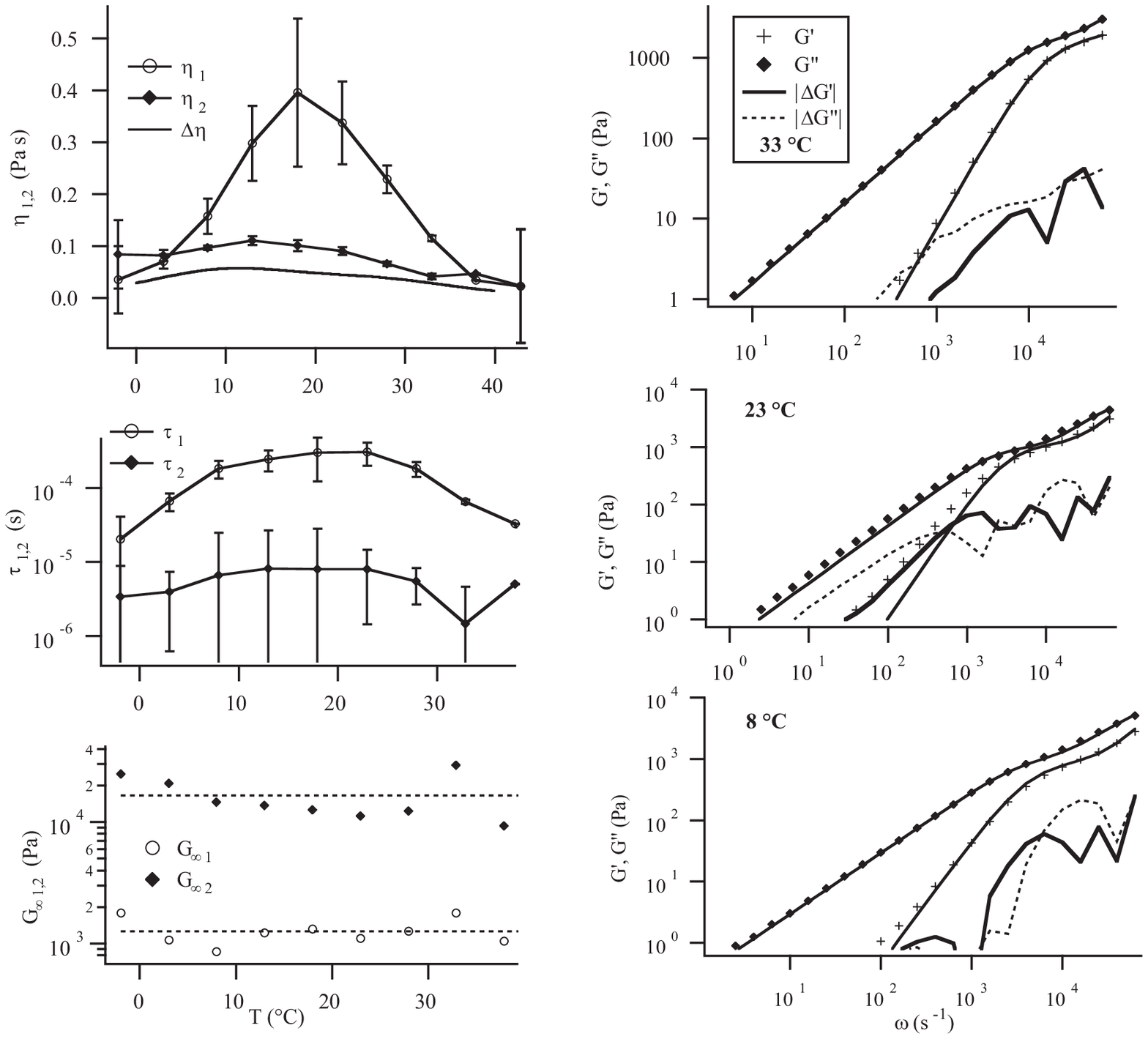}}
\caption{\protect\small Rheology data for the {\formeau} system at
$c=35 \%$. Left~: Value of the fit parameters in equation
(\ref{eq:2maxwell}), as well as $G_{\infty i} = \eta_{i} /
\tau_{i}$. The upper graph also shows the difference between the
viscosity measured by Strey \cite{strey96} (at $\omega = 0$) and
our values at $\omega = 600 \, \rm{s}^{-1}$ ($\Delta \eta = \eta
(0)-\eta(600)$). Right~: The value of $G'$ and $G''$ as a function
of $\omega$ and the corresponding fit with two Maxwell models (for
$33 \dgr$, $23 \dgr$ and $8 \dgr$). Thin line~: fit to the data;
thick line~: residue of $G'$; dotted line~: residue of $G''$.}
\label{fig:data35}
\end{figure}

\noindent We performed rheology measurements in the isotropic
phase of {\formeau} at different concentrations and temperatures.
We present in figure \ref{fig:eta_total} the values of the
viscosity at $\nu = 100 \un{Hz}$ or $\omega \simeq 600
\un{s^{-1}}\,$ \footnote{At lower frequency, the data becomes
noisy (especially at low concentrations), but there is no
systematic variation in viscosity with respect to $\omega = 600
\un{s^{-1}}\,$.}, as well as the values obtained by Strey
\cite{strey96} using capillary rheometry (at zero frequency) for
$c \in \{ 5,10,18,25,35 \} \, \%$, shown in full or dotted
line. Note that at high concentration ($c \geq 50 \%$ for
{\formeau} and $c \geq 30 \%$ for
\mbox{C$_{12}$EO$_{8}$/D$_{2}$O$\,$}) the curves stop abruptly at
lower temperature because of the presence of the mesophases (see
reference \cite{mitchell83} for the phase diagrams) and of a
crystalline phase for the pure surfactants. Our results are in
good agreement with those of Strey, who employed a totally
different technique. This validates the precision of our technique
and confirms that sample concentration does not drift during the
experiment. The difference $\Delta \eta = \eta (0)-\eta(600)$
between the values of Strey (at $\omega = 0$) and our results for
$\omega = 600 \, \rm{s}^{-1}$ could be due to lower frequency
relaxation mechanisms or, more probably, it only reflects the
uncertainty of the measurement. We have howevered plotted $\Delta
\eta$ for each concentration in figures \ref{fig:data35} and
\ref{fig:data1825}.

The low frequency viscosity is the sum of the viscosities
associated to the various relaxation mechanisms. Performing
rheology measurements over a wide frequency range allows us to
separate those mechanisms. We now present the frequency behavior
of $G^* (\omega )$ for the different concentrations.

\begin{figure}[htbp]
\centerline{\includegraphics[width=17.5cm]{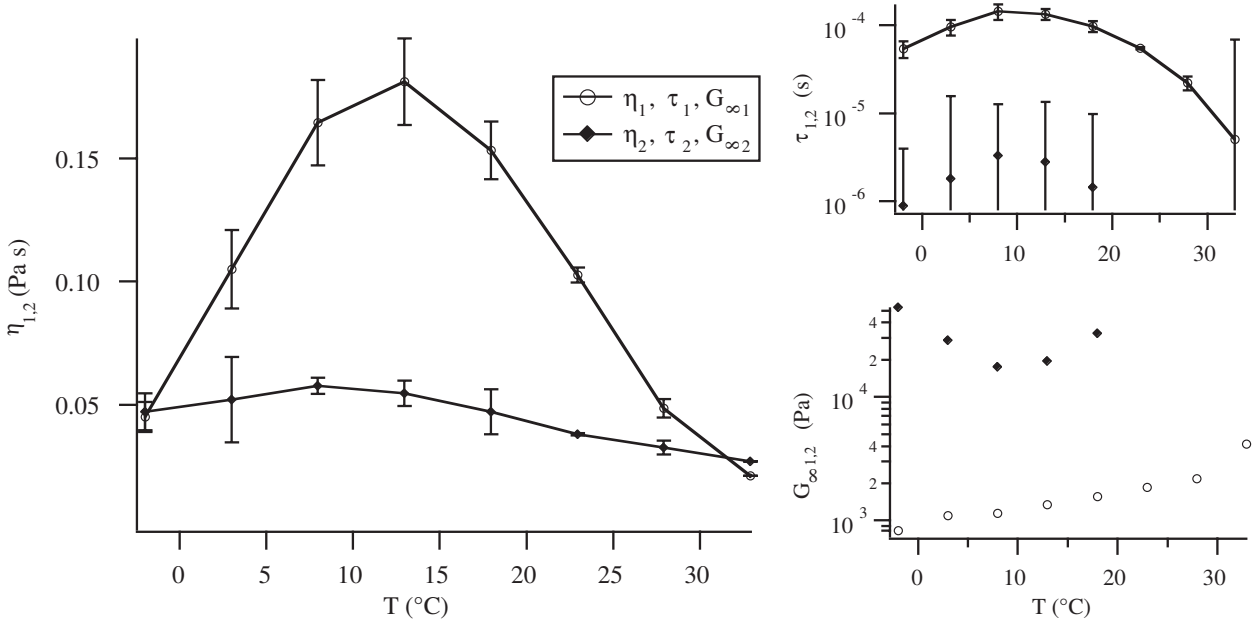}}
\caption{\protect\small Rheology data for the {\formeau} system at
$c=30 \%$.} \label{fig:data30}
\end{figure}

\begin{figure}[htbp]
\centerline{\includegraphics[width=17.5cm]{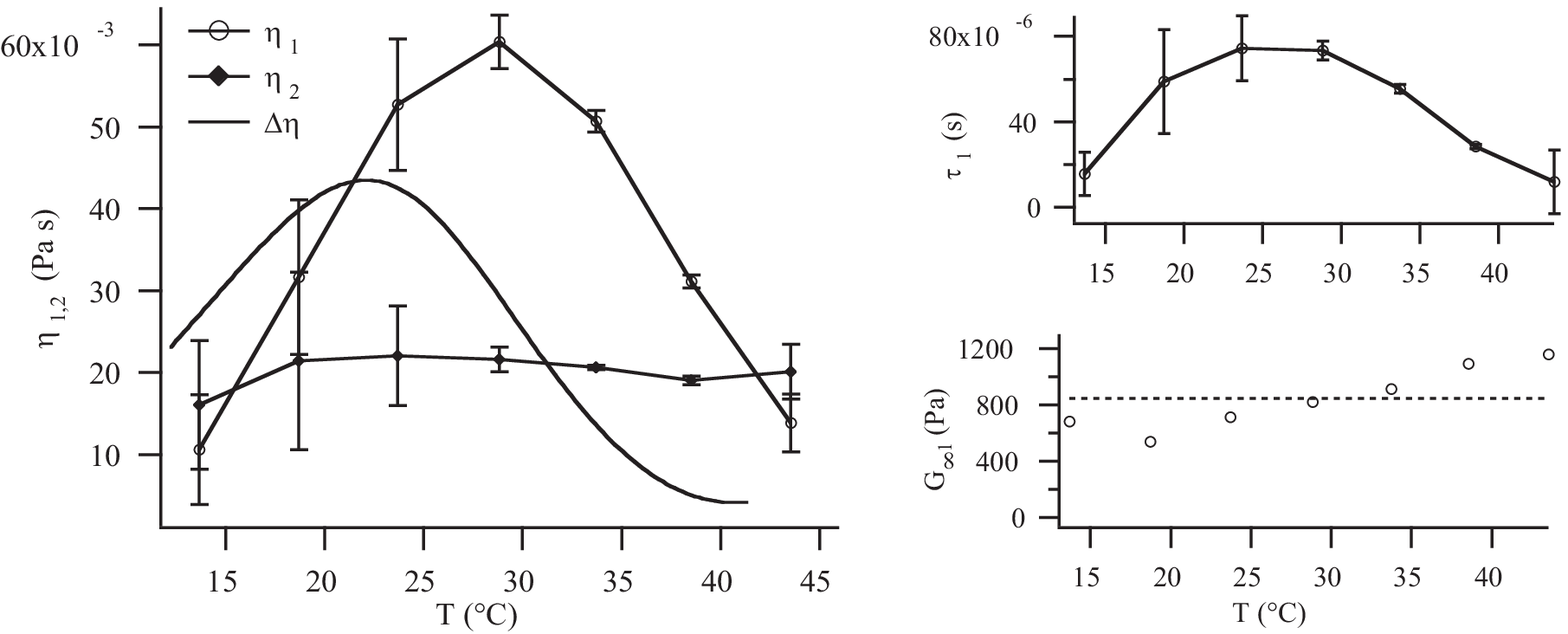}}
\centerline{\bf a)}
\centerline{\includegraphics[width=17.5cm]{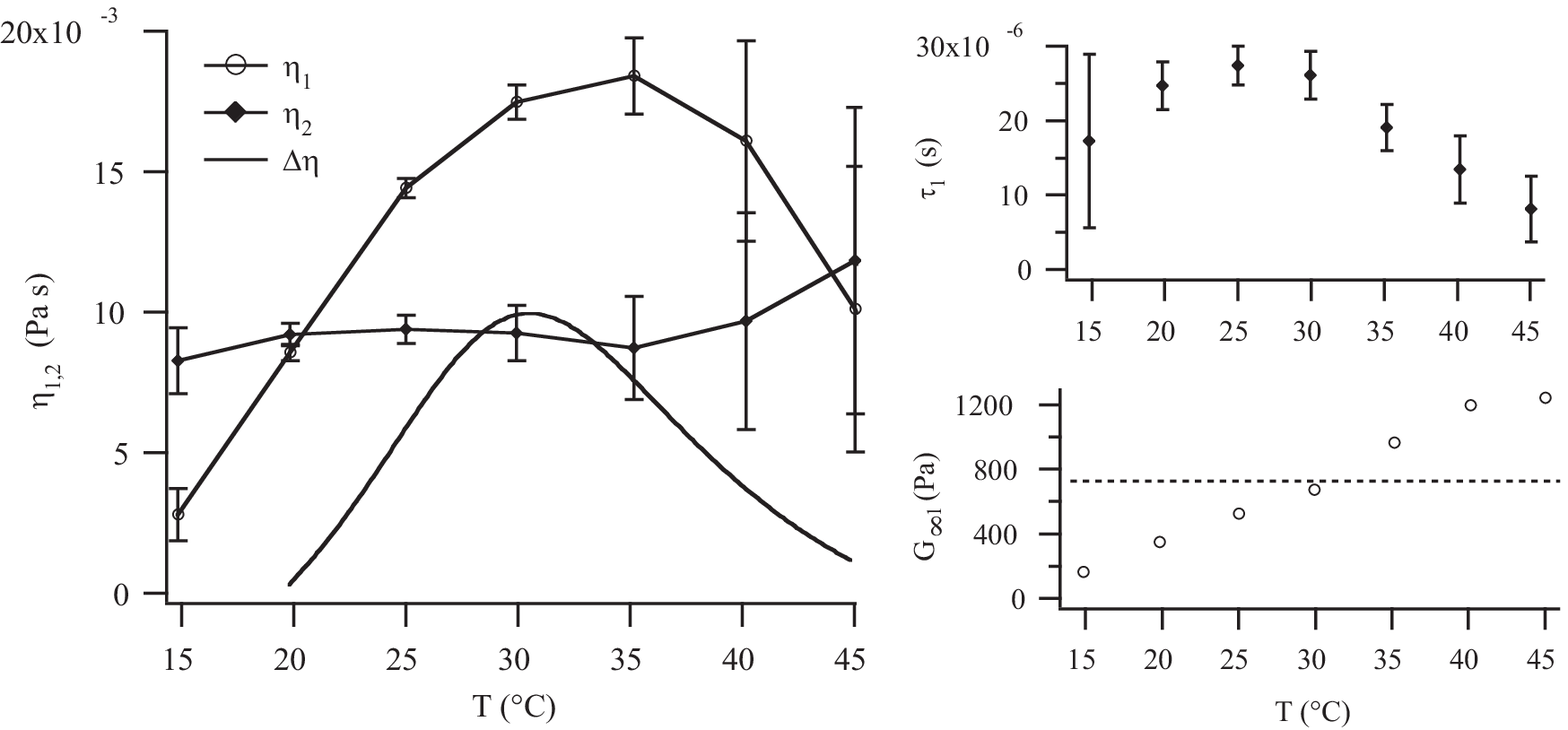}}
\centerline{\bf b)} \caption{\protect\small Rheology data for the
{\formeau} system at $c=25 \%$ a) and $18 \%$ b).}
\label{fig:data1825}
\end{figure}

For higher concentrations ($c=35$ and $30 \%$), the curves are
well fitted with the sum of two Maxwell models
(\ref{eq:2maxwell}). In figure \ref{fig:data35} we show the data
for $c=35 \%$~: $G'$ and $G''$ as a function of $\omega$ for three
temperature points ($8$, $23$ and $33 \dgr$) and the value of the
fit parameters in equation (\ref{eq:2maxwell}).
Equation (\ref{eq:2maxwell}) adequately describes the data for
$c=30 \%$, too. Figure \ref{fig:data30} shows the fit parameters.
Note that ${\tau}_2$ is smaller (and its values are less reliable)
than for $c=35 \%$.
For $c=18$ and $25 \%$ the fit with equation (\ref{eq:2maxwell})
yields a very small ${\tau}_2$ and fit quality does not change if
we set ${\tau}_2=0$ (which amounts to adding a constant viscosity
background to the first Maxwell model). The fit parameters are
represented in figure \ref{fig:data1825}.

Finally, for the lower concentrations ($c=5$ and $10 \%$), no
storage modulus $G'$ is detected. The solutions exhibit simple
viscous behavior and the viscosity as a function of temperature is
shown in figure \ref{fig:eta_total}.

On the other hand, there is no sign of the slow relaxation mode
for the \mbox{C$_{12}$EO$_{8}$/D$_{2}$O$\,$} system; for $c \leq
25 \%$ the system is purely viscous in the investigated frequency
range (there is no detectable storage modulus), and the values for
the low-frequency viscosity are given in figure
\ref{fig:eta_total} (right).

\begin{figure}[htbp]
\centerline{\includegraphics{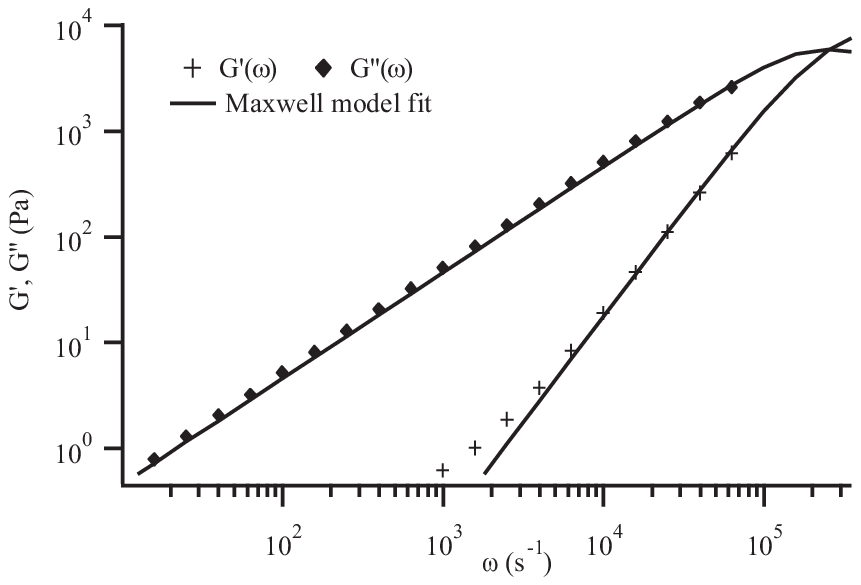}} \centerline{\bf a)}
\centerline{\includegraphics{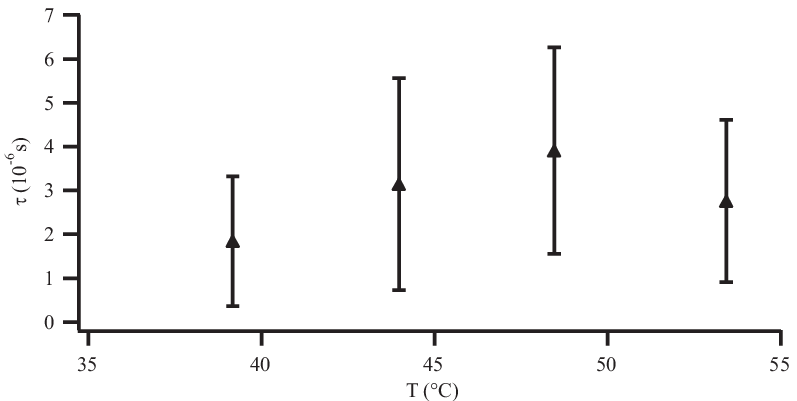}} \centerline{\bf b)}
\centerline{\includegraphics{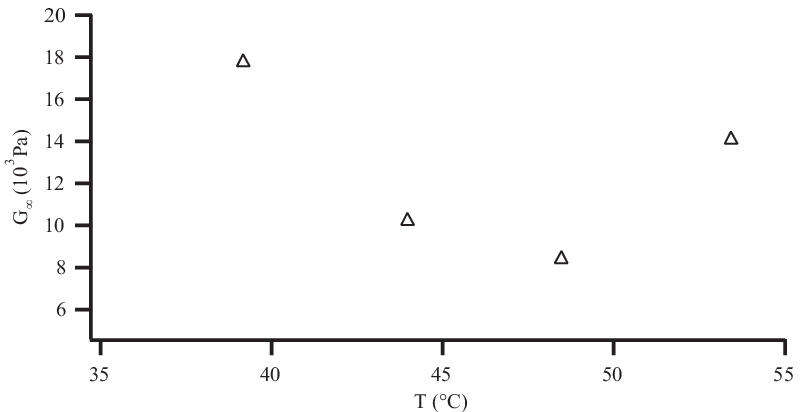}} \centerline{\bf c)}
\caption{\protect\small Rheology results for the
\mbox{C$_{12}$EO$_{8}$/D$_{2}$O$\,$} mixture at $c=30 \%$~: a)
Maxwell model fit at $T=48.5 \dgr$ b) relaxation time $\tau$ and
c) plateau modulus $G_{\infty}$ as a function of temperature.}
\label{fig:tau30}
\end{figure}

In accordance with previous results \cite{constantin02}, we assign the fast mechanism in the {\formeau} to the
relaxation of local order; both the relaxation time $\tau _{2}$ (in the $\un{\mu s}$ range) and the plateau modulus
$G_{\infty 2} \sim 2 \, 10^4 \un{Pa}$ at $30-35 \%$ -- figures \ref{fig:data35} and \ref{fig:data30}) are coherent
with the values previously obtained for $c=50 \%$. This mechanism is also detected in the
\mbox{C$_{12}$EO$_{8}$/D$_{2}$O$\,$} system at $c=30 \%$; the corresponding viscosity is presented in figure
\ref{fig:eta_total} (right) and the relaxation time  and plateau modulus (only detectable by our technique between
about $40-55 \dgr$) are given in figure \ref{fig:tau30}.

As to the slow relaxation mechanism, it is only present in the {\formeau} system, with longer micelles, and it does not
appear in the {\mbox{C$_{12}$EO$_{8}$/D$_{2}$O$\,$}} solution, where the micelles are shorter. We can therefore
consider that it is due to the entangled network relaxing by micellar reptation. The relaxation time $\tau _{1}$ first
increases with temperature, goes through a maximum and subsequently decreases. The viscosity $\eta _{1} = G_{\infty 1}
\tau _{1}$ has a similar evolution. The temperature position of the maximum goes from $18 \dgr$ at $c=35 \%$ to $35
\dgr$ at $c=18 \%$. As discussed in the theoretical section, this variation can be understood as an increase in
micellar size, followed by the appearance of connections. This variation is coherent with the fact that the curvature
of the aggregates diminishes with increasing temperature (due to the decreasing hydration of the nonionic polar groups
\cite{puvvada90,israelachvili92}), favoring low-curvature junctions over high-curvature end-caps. $\tau _{1}$ takes
values around $ 10^{-4} \un{s}$ (figures \ref{fig:data35}-\ref{fig:data1825}), much smaller than the usual reptation
times in wormlike micellar systems \cite{aitali97,narayanan97,khatory93b}; the difference could be a sign that the
entangled micelles become connected before they reach a sizeable length. An alternative explanation for this fast
dynamics would be a very short breaking time $\tau _{\rm{br}}$.

\begin{figure}[htbp]
\centerline{\includegraphics{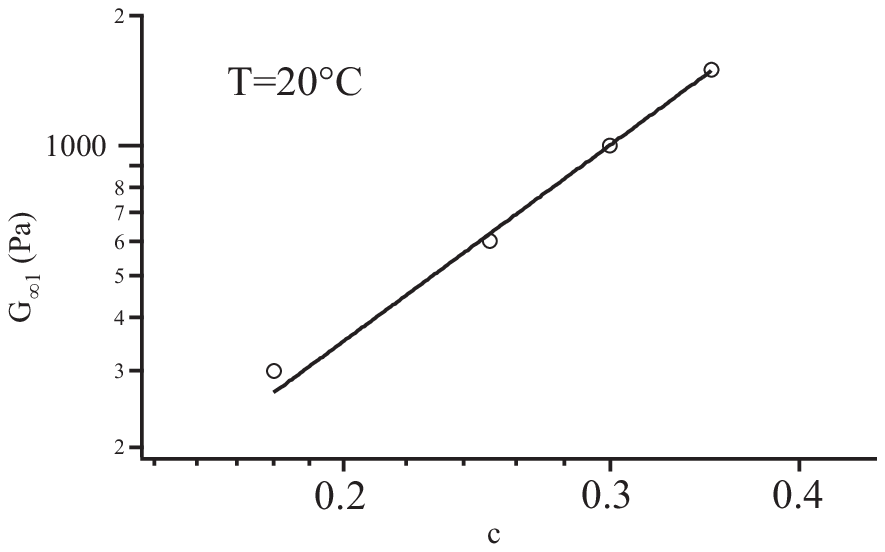}} \caption{\protect\small
Plateau modulus $G_{\infty 1}$ of the slow relaxation mode in the
{\formeau} system at $20 \dgr$ as a function of the concentration.
The line is a power-law fit with exponent $2.6$.} \label{fig:gofc}
\end{figure}

The corresponding plateau modulus $G_{\infty 1}$ changes both with the concentration and the temperature. Especially
at the lower concentrations (figure \ref{fig:data1825}), $G_{\infty 1}$ increases with temperature, a variation which
might be related to the appearance of connections or to a change in persistence length. Let us consider in more detail
the concentration dependence of $G_{\infty 1}$ in the low-temperature ($15-20 \dgr$) range, where the micelles are not
yet connected. As shown in figure \ref{fig:gofc}, its variation can be described by a power law~: $G_{\infty 1}
\propto c^{2.6 \pm 0.1}$. For semidilute polymer solutions, this exponent can be theoretically estimated at $9/4$ but
is actually slightly higher, depending on the quality of the solvent \cite{doi86}. This result further reinforces our
conclusion that the slower mode is due to the relaxation of the entangled network.

We would like to point out that solutions of lecithin/$\un{H_2O}$/$n$-decane exhibit a similar bimodal relaxation
spectrum (as recently shown by Schipunov and Hoffmann \cite{shchipunov98}), where the low-frequency mechanism seems to
disappear upon heating (see their figure 18). Since these systems are very viscous and the entire relaxation spectrum is
accessible to conventional rheometers, a systematic investigation of their behavior in concentration and temperature
could yield interesting information, as Laplace transform techniques could be used on the rheology data in order to
obtain the distribution of relaxation times $H(\tau)$ \cite{ferry80}. Unfortunately, we cannot apply this technique
for our system because our frequency range does not cover the entire relaxation spectrum. This is why we limited
ourselves to the simplest phenomenological model (two Maxwell relaxations), which is nevertheless sufficient for
separating the contribution of the two physically relevant mechanisms.

\section{Conclusion}

In the present work, we use high-frequency rheology to elucidate
the dynamical properties of the isotropic phase in the nonionic
surfactant system {\formeau}. We show that the relaxation spectrum
can be interpreted as the sum of two mechanisms; one of them, with
a relaxation time in the microsecond range, varies little in
temperature and can be attributed to the relaxation in local
micellar order. The second one, at about $ 10^{-4} \un{s}$,
corresponds to the relaxation of the entangled network and its
temperature variation (increase, then decrease) can be explained
by the micelles lengthening and then becoming connected. When the
system is highly connected, only the fast mode --related to the
local order-- is significant. The slower mode does not appear in
the \mbox{C$_{12}$EO$_{8}$/D$_{2}$O$\,$} system, where the
micelles are much shorter, confirming that it is related to the
existence of the entangled network.

We conclude that our results show the importance of local order in
concentrated isotropic surfactant phases; local order relaxation
can even dominate the dynamical behaviour of the system when
slower mechanisms (reptation etc.) are irrelevant. We also hope to
illustrate the usefulness of high-frequency rheology for the study
of surfactant solutions.

\end{document}